# Financial markets as a Le Bonian crowd during boom-and-bust episodes: A complementary theoretical framework in behavioural finance

*Abstract: This article proposes a complementary theoretical framework in behavioural finance by interpreting financial markets during boom-and-bust episodes as a Le Bonian crowd. While behavioural finance has documented the limits of individual rationality through biases and heuristics, these contributions remain primarily microeconomic. A second, more macroeconomic strand appears to treat market instability as the aggregated result of individual biases, although it generally does so without an explicit theoretical account of how such aggregation operates. In contrast, this paper adopts a macro-psychological — and therefore macroeconomic — perspective, drawing on classical crowd psychology (Le Bon, 1895; Tarde, 1901; Freud, 1921).*

*The central claim is that during speculative booms and crashes, markets behave as psychological crowds governed by unconscious processes, suggestion, emotional contagion, and impulsive action. These episodes cannot be understood merely as the sum of individual departures from rationality, but as the emergence of a collective mental state that follows its own psychological laws.*

*By reintroducing crowd psychology into behavioural finance, this paper clarifies the mechanisms through which market-wide irrationality arises and offers a theoretical foundation for a macro-behavioural understanding of financial instability.*

Claire Barraud[1]

## Introduction

If one were to ask: *What image spontaneously comes to mind when you hear the expression "speculative bubble" or "financial crash"?* Perhaps the immediate associations are those of a human stampede, a tidal wave, a spreading wildfire, a frenzied crowd, or a panicked multitude. Many of the dominant metaphors mobilised to describe financial crises are, in fact, crowd metaphors (Canetti, 1962).

Sigmund Freud famously suggested that there are ultimately only two types of sciences: the natural sciences and psychology (Le Guen, 2022). Economics appears, at first glance, to occupy an intermediate space between the two, sometimes drawing its conceptual inspiration from the natural sciences, sometimes from psychology, and increasingly from both at once. In *Narrative Economics: How Stories Go Viral and Drive Major Economic Events*, Robert Shiller (2019) reintroduces what William Whewell (1840) called the challenge of consilience—that is, "*the unity of knowledge among the differing academic disciplines, especially between the sciences and the humanities"* (Shiller, 2019, p. 12).

The rise of behavioral finance has made it possible for economics to internalize insights from psychology. This development marked a revolution in thinking for economic thought beginning in the late 1970s. Its conceptual seeds can be traced back to Keynes (1936), through his famous "beauty contest" and the notion of differentiated degrees of rationality, and later to Simon (1955) with the idea of bounded rationality. Behavioral finance subsequently expanded and empirically substantiated these intuitions. From this perspective, economics can be understood as the study of human psychology in transactional settings, which, despite its particular analytical rigor and methodological specificity, could be viewed as a branch of psychology in its own right.
Since these pioneering contributions, “the results obtained by analysing the 1987–2017 period show a growth potential of Behavioural Finance. Investor sentiment is the main subject among the thirteen main subjects of this area” (Paule-Vianez et al., 2020, p. 71). For my part, it is the sentiment of the market as a whole that underpins my analytical approach.

[1] PhD in Economics and teacher, Grenoble Alpes University, France.

This article is situated within the broader tradition of behavioral finance, while proposing a complementary — and in some respects broader — analytical framework by adopting a macroeconomic perspective grounded in the tools of crowd psychology developed by Gustave Le Bon (1895) and later revisited by Sigmund Freud ([1921] 2014)[2]. The aim is to shed new light on the functioning of financial markets under extreme conditions, that is, during speculative bubbles and financial crises.
These very circumstances are also those from which behavioral finance itself emerged. As Serge Moscovici (1985, p. 35) observed, “for a science to be born, it is not enough for a phenomenon to exist — it has been known for thousands of years. Nor is it sufficient that its strangeness merely fascinates a few inquisitive minds. It must proliferate to such an extent that, from something episodic or harmless, it becomes pervasive, disruptive, and an object that can no longer be ignored.” What some economists initially regarded as a mere “anomaly” has, since the wave of financial liberalization, become so frequent and persistent that it has turned into a central object of inquiry: the extreme volatility of financial market prices.

The neoclassical tradition, from its inception in the 1870s, sought to construct *homo economicus* on the basis of utilitarian philosophical theory. According to this view, individuals are driven by an irrepressible need to maximise pleasure (i.e., utility) and minimise pain (i.e., disutility). The neoclassical *homo economicus* is assumed to be perfectly rational—not only in the sense that the means employed are consistent with the ends pursued, but also in the sense that the chosen solution is always optimal given the constraints of the environment. It is within this logic of substantive rationality that the efficient market hypothesis (EMH, henceforth) emerged, beginning with Eugene Fama’s first formulation in 1965. In financial markets, the representative economic agent is assumed to form rational expectations: selecting the optimal investment by examining the issuer’s fundamentals, and doing so in a probabilistic universe in which expected utility theory à la Markowitz applies. Market participants are therefore conceived as fundamentalists, relying on informationally grounded rationality to construct asset portfolios according to an optimal risk–return trade-off consistent with their preferences. Fama (1970) did acknowledge the existence of *noise traders*, but—like Friedman (1953)—did not consider them capable of destabilising prices. Because prices are presumed to remain close to intrinsic value, irrational participants, who buy as prices rise and sell as they fall, are deemed ultimately inconsequential: market forces will either eliminate them through accumulated losses or drown them out, as rational traders—the presumed majority—counteract their sporadically irrational actions.

The persistent, significant, and recurrent divergences between market prices and the intrinsic value of assets led a number of scholars to question the efficient market hypothesis and, by extension, the capacity of financial markets to guarantee macroeconomic stability. Behavioral finance emerged in the wake of the pioneering work of Tversky and Kahneman (1974, 1979), who challenged the relevance of the Markowitz framework by developing what would later become the famous Prospect Theory. In essence, everything is a matter of perspective — including rationality itself. Within this framework, as in many theories in psychology, *objective reality* does not operate as such; instead, judgment is filtered through subjective perception, which is framing-dependent and therefore malleable.

[2] The quotations from Le Bon (1895), Moscovici (1985), Tarde (1901) are translated by the author from the original French editions, after cross-checking with the available English versions for coherence.

Building on Simon's notion of bounded rationality, Kahneman and Tversky demonstrated that human rationality is both partial and biased, systematically guided by heuristics — routinised cognitive shortcuts that have become automatic. Simon had explained this in terms of procedural rationality, according to which agents follow a heuristic until they reach a "satisficing" rather than optimal outcome, because doing so economises on cognitive costs and information-processing. Kahneman and Tversky reached a different conclusion: heuristics are not merely economising devices but intrinsic mental habits, embedded in the architecture of human cognition.

Since then, behavioral finance has continued to expand, identifying and repeatedly validating a wide range of cognitive biases inherent to human decision-making. Today, at least a dozen such biases are widely recognised. These biases correspond, in my view, to the *microeconomic* dimension of behavioral finance; and one may legitimately infer that their aggregate expression at the collective level makes *macroeconomic* behaviour equally prone to deviation. Moreover, these biases are characteristic of System 1 of thinking, as described by Stanovich and West (2000) and later by Kahneman (2002; 2012). Crucially, crowds operate almost exclusively through System 1.

In parallel, other scholars developed a more explicitly *macroeconomic* approach to behavioral finance, often referring — implicitly or explicitly — to financial markets as a "crowd." But the market as a whole is most often understood as the aggregation of biased rationalities. Robert Shiller has been one of the major contributors to this perspective since the early days of behavioral finance, culminating in his 2000 book *Irrational Exuberance*, which synthesises this line of inquiry. His work stands in continuity with major historians of economic phenomena such as Charles Kindleberger (1978), whose aptly titled *Manias, Panics and Crashes: A History of Financial Crises* offered a historical genealogy of speculative euphoria and collapse, and Hyman Minsky, whose "financial instability hypothesis" (1977) later led to the formulation of the "Paradox of tranquillity" (1986).

These two analytical perspectives within behavioral finance, which constitute the focus of the first part of this article, will then serve as a foundation for the second part, where we zoom out even further and enter the domain of crowd psychology. Crowd psychology is a branch of social psychology. Social psychology is the field that studies "the way in which the thoughts, feelings, and behavior of individuals are influenced by the actual, imagined, or implied presence of others" (Allport, 1985, p. 3). Crowd psychology goes beyond this definition by treating the crowd as a self-contained entity distinct from the individuals who compose it, and by placing central emphasis on the role of the leader. As Moscovici observes, "crowd psychology was born the day its pioneers began to ask the question that was on everyone's lips: how do leaders exercise such extraordinary power over the masses?" (Moscovici, op. cit., p. 20).

The major advances of behavioral finance have significantly improved our understanding of financial markets, whose intrinsic instability gives rise to the exceptional circumstances of speculative bubbles and crashes. The added contribution of this article lies in the additional analytical shift made possible by crowd psychology. First, Le Bon's notion of the "psychological crowd" (op. cit., p. 16) emerges under precisely these extraordinary circumstances. Second, we show that the interpretive framework offered by crowd psychology closely parallels that of behavioral finance, particularly in its macroeconomic orientation. And finally — perhaps most

importantly — crowd psychology provides new conceptual tools for understanding how markets operate during episodes of extreme stress. It can therefore enrich behavioral finance by extending it into an explicitly macroeconomic and macropsychological undertaking.
Because as Le Bon reminds us, "knowledge of crowd psychology is the indispensable resource of the statesman who seeks, not to govern crowds — a task that has today become exceedingly difficult — but at least not to be governed by them" (op. cit., p. 13).

# 1. Behavioral Finance: Microeconomic Foundations and Macroeconomic Extensions

In order to address our central research question — namely, whether financial markets can be understood as *Le Bon–style* crowds — we distinguish between two strands of foundational work in behavioral finance: a microeconomic (i.e. micropsychological) approach and a macroeconomic (i.e. macropsychological) approach. The microeconomic approach has been primarily concerned with identifying individual-level departures from rationality. The macroeconomic approach, by contrast, examines financial markets either as a collective entity in their own right or as the aggregate manifestation of underlying cognitive biases (for a synthesis, see Table 2, p. 15.)
Building on this distinction, our analysis will subsequently focus on the macroeconomic approaches, in order to compare the functioning of financial markets as described by economists with the defining features of crowds in Gustave Le Bon's framework.

## 1.1. From Individual Bounded Rationality : Cognitive Biases

Behavioral finance thus challenges the core assumptions underlying market efficiency, namely the rationality of agents and the idea that rational investors — or the market mechanism itself — can neutralise or eliminate the actions of noise traders.
The pioneering work of Kahneman and Tversky in cognitive psychology shed light on the various cognitive biases to which all individuals are susceptible, particularly under conditions of risk or uncertainty. Prospect Theory offers an alternative to the expected utility framework of von Neumann and Morgenstern (1944), to Bayesian inference, and to Markowitz's model. It demonstrates that we are not good statisticians: when faced with uncertainty or risk, our judgments, decisions, and actions are systematically shaped by the *framing* that induces a given individual perspective. As Thaler (1985, p. 201) puts it, the objective is "to replace the utility function from economic theory with the psychologically richer value function." If all economic choice is about *perceived* value, then all decision-making is ultimately governed by individual prospect.

***The law of small numbers***
Kahneman and Tversky (1974), building on their earlier demonstration of the "law of small numbers" (Kahneman & Tversky, 1971) and the sampling errors that follow from it, showed that under conditions of risk individuals typically exhibit the *certainty effect*, preferring a sure gain to a probabilistic one. Conversely, according to the *reflection effect*, they become *risk seeking* in the domain of losses, preferring to gamble on a probable loss rather than accept a smaller but certain one. In financial markets, these two tendencies were empirically documented by Shefrin and Statman (1984), who showed that investors tend to sell too quickly when facing gains and hold on to losing positions for too long. The authors already acknowledged the emotional dimension of this mechanism, since the *disposition effect* in the loss domain is partly driven by *regret aversion*. Their approach also drew on several aspects of Thaler's (1985) work. First, *mental accounting* highlights the fact that individuals segment gains and losses asymmetrically, with their evaluations strongly shaped by recent outcomes. Second, the *endowment effect* also helps to explain conservatism in the loss domain: individuals overvalue what they already own when faced with the possibility of selling, and are therefore reluctant to accept a "fire-sale" price. In 1993, Thaler and Benartzi developed the concept of *myopic loss aversion*, grounded in the observation that investors tend to have *short evaluation horizons*, which leads them to demand a high risk premium in order to compensate for short-term losses, even at the expense of long-term profitability (Benartzi & Thaler, 1993). The *myopic loss aversion* framework thus resolves the *equity premium puzzle* (Mehra & Prescott, 1985) by explaining why equities *outperformed comparatively to bonds*.

Within the 1979 Prospect Theory framework, Kahneman and Tversky devoted significant effort to demonstrating the *framing effect*, which itself gives rise to three distinct cognitive biases.

**The r*epresentativeness bias***
First, *representativeness bias* leads individuals to reason on the basis of a limited subset of information — typically the most salient or immediately accessible — and thereby induces the *law of small numbers*, which causes people to overestimate the reliability of small samples. In financial markets, this bias was documented by Barberis, Shleiffer, and Vishny (1997) as a source of underreaction to news when conservatism prevails, and a source of overreaction *to good and bad news* when decision-making relies on a narrow set of observations rather than on the *laws of probability* over the long run.

**The a*vailability bias***
Second, *availability bias* leads individuals to overweight information that is most readily retrievable from memory, particularly when such information is stereotyped, vivid, or emotionally charged.

**The s*ubstitution bias***
Finally, *substitution bias* leads individuals to answer a difficult question by unconsciously substituting it with a simpler one, thus producing a response that is at best partial, and at worst beside the point.

**[Freeze frame:] *The two systems of thinking***
To synthesise the advances of Prospect Theory, Kahneman later formalised what had been implicit in his earlier work with Amos Tversky by referring to these two modes of cognition as "systems" (Kahneman, 2002). From the outset — already in 1971 with the *law of small numbers* and throughout their subsequent research — Kahneman and Tversky distinguished intuition from reasoning, contrasting the full rationality of *homo economicus* with the truncated judgment of the average decision-maker, by comparing what one *should* do under complete rational deliberation with what one actually tends to do instinctively and with little conscious control.
The first authors to explicitly use the terminology of *System 1* and *System 2* were Keith Stanovich and Richard West (2000, p. 658), who define System 1 "as automatic, largely unconscious, and relatively undemanding of computational capacity. Thus, it conjoins properties of automaticity and heuristic processing as these constructs have been variously discussed in the literature." In 2002, Kahneman adopted these terms to distinguish intuitive from deliberative reasoning.
Kahneman (op. cit. p. 698) further explains that "the operations of System 1 are typically fast, automatic, effortless, associative, implicit (not available to introspection), and often emotionally charged; they are also governed by habit and are therefore difficult to control or modify. The operations of System 2 are slower, serial, effortful, more likely to be consciously monitored and deliberately controlled; they are also relatively flexible and potentially rule governed". Moreover, "the perceptual system and the intuitive operations of System 1 generate impressions of the attributes of objects of perception and thought. These impressions are neither voluntary nor verbally explicit. In contrast, judgments are always intentional and explicit even when they are not overtly expressed. Thus, System 2 is involved in all judgments, whether they originate in impressions or in deliberate reasoning" (p. 699). However, System 2 does not necessarily succeed in correcting the errors of System 1, because "the [System 2] monitoring is normally quite lax and allows many intuitive judgments to be expressed, including some that are erroneous" (p. 699).

Accordingly, it is unsurprising that most cognitive biases belong to System 1 and persist even when System 2 is available in principle to override them (see Table 1, p. 8). The only major exception is *confirmation bias*, which can be classified as a System 2 process, insofar as it involves a deliberate — albeit selective — search for evidence that corroborates one's prior beliefs, even though one may also plausibly argue that the bias originates in System 1 when no corrective intervention from System 2 occurs.

It is worth noting at this point that Kahneman's distinction — within the strictly cognitive domain — parallels what psychoanalysts since Freud have identified in a broader conception of the human psyche: the coexistence of consciousness and the unconscious. In psychoanalytic theory as well, it is the unconscious that governs without the awareness of the conscious mind, such that "the ego [i.e. consciousness] not master in its own house" (Freud, 1933).

In short, Kahneman's framework, and that of his various co-authors, remains one of individual rationality under conditions of uncertainty or risk.

Following the work of Kahneman and Tversky, many scholars in both economics and psychology turned their attention to the biases that hinder rational decision-making at the individual level, or, in our terms, at the microeconomic level. In a manner still rather unflattering for us human beings—so convinced of our own capacity to reason at least somewhat correctly—other biases were subsequently identified or rediscovered.

***The insight bias***

Fischhoff (1975) cautioned us against the insight bias, our tendency to rationalise ex post events that, in most cases, made little sense to begin with. This bias leads us to believe, after the fact, that an event was more predictable than it actually was, sometimes even when it was not predictable at all.

***The self-attribution bias***

Building on Heider's (1958) distinction between internal (dispositional) and external (situational) causal attributions, Miller and Ross (1975), followed by Daniel, Hirshleifer and Subrahmanyam (1998), developed the concept of self-attribution bias. This bias leads us to attribute successes to internal disposition while ascribing failures to external causes. Miller and Ross refer to these as *self-serving biases* to highlight, rather bluntly, the egocentric nature of this tendency. Daniel, Hirshleifer and Subrahmanyam then tested this mechanism empirically and concluded that it is a driver of both underreaction and overreaction in financial markets.

***The overconfidence bias***

The self-attribution bias is directly responsible for the overconfidence bias, also discussed by Miller and Ross (1975), through what they call a self-enhancing effect, which leads individuals to expect their own behaviour to result in success. At the aggregate level, this bias also generates overreactions in financial markets, as shown by Daniel, Hirshleifer and Subrahmanyam (op. cit.). Its empirical confirmation can be traced to Fischoff, Slovic and Lichtenstein (1977, p. 1), who observed it "across several different question and response," to the point that subjects were "willing to stake money." In financial market settings, the emotional dimension of decision-making was further documented by Lo and Repin (2002), who measured physiological stress responses in ten traders exposed to what they termed “transient market events,” thereby demonstrating the magnitude of emotional involvement (p. 1).

***The affect bias***

In parallel, the affect bias, or affect heuristic, is among the most extensively studied cognitive biases. First identified by Zajonc (1980), it prevents individuals from evaluating events and behaviours purely rationally. According to Zajonc, affective preferences shape our evaluations and choices even before any analytical reasoning takes place. Slovic (1987) further showed that risk assessments are guided less by rational deliberation than by affective reactions such as fear and sympathy. It was not until 2000 that Slovic and his colleagues formally named this mechanism the affect heuristic, describing it as a bias in risk perception operating especially under time pressure (Finucane, Alhakami, Slovic and Johnson, 2000). Kahneman and Frederick (2002) later argued that affect should replace anchoring as the primary explanatory mechanism for the persistence of first impressions in judgment.

***The mimetic bias***

Finally, well known since the work of Keynes in macroeconomics and finance (1936), and operating at both the individual and collective level (group bias), the mimetic bias arises when individuals copy the beliefs and behaviours of others instead of relying on their own information. This bias, which manifests as *herding* in financial markets, was first described as a general property of living beings by Gabriel Tarde in *Les lois de l'imitation* (*The laws of imitation*, 1901). According to Tarde, all living organisms imitate their environment in order to survive, adapt, and learn. René Girard (1961) later developed his theory of mimetic desire, arguing that our desires are not determined by the intrinsic qualities of the object itself, but by the desire that others project onto it. Quite often, without realising it, we desire what others desire.

In the financial domain, Keynes characterised mimetic behaviour as a *rational* strategy. Faced with radical uncertainty, he argued that it is safer for one's reputation to follow the crowd — *the convention* — than to stand alone, recalling that “worldly wisdom teaches that it is better for reputation to fail conventionally than to succeed unconventionally” (Keynes, 1936, pp. 100–101). For Keynes-as-speculator, rationality in financial markets — given their high degree of uncertainty and the objective of “beat the market” — consists, at best, in anticipating what others will do, and at worst, in imitating them, that is, following the crowd.

André Orléan (1986) subsequently extended Keynes's notion of convention to show that it coordinates and guides behaviour in situations of uncertainty, thereby theorising the mechanisms of mimetic interaction that prevail in financial markets. Finally, Alan Kirman (1993) reminded economists that mimetic behaviour is inherent to all living beings by drawing on the recruitment patterns of ants choosing between alternative food sources, arguing that human beings are no different in this respect — including in market exchanges.

**The resulting noise bias**

Taken together, these biases — even considered individually — generate a level of “noise” that makes supposedly objective decision-making highly variable across individuals (Kahneman, Sibony and Sunstein, 2021).

**Table 1: Categorization of Cognitive Biases by Kahneman's System 1 and System 2**

**System 1 Biases (intuitive, fast, automatic thinking)**

| **Bias / Heuristic** | **First author(s)** | **Definition** |
|---|---|---|
| Law of Small Numbers | Kahneman & Tversky (1971) | Overestimating how well small samples represent the population. |
| Representativeness Bias | Kahneman & Tversky (1972) | Judging probability based on similarity to a stereotype. |
| Availability Bias | Tversky & Kahneman (1973) | Estimating frequency based on ease of recalling examples. |
| Anchoring Bias | Tversky & Kahneman (1974) | Judgment influenced by an initial reference value (anchor). |
| Substitution Bias | Tversky & Kahneman (1974); Kahneman & Frederick (2002) | Replacing a complex question with a simpler one. |
| Hindsight Bias | Fischhoff (1975) | Believing, after the fact, that an event was more predictable than it was. |
| Self-Attribution Bias | Heider (1958); Miller & Ross (1975); Daniel et al. (1998) | Attributing successes to oneself and failures to external causes. |
| Overconfidence Bias | Fischhoff et al. (1977); Alpert & Raiffa (1977) Kahneman et al. (1982) | Overestimating the accuracy of one's judgments. |
| Affect Heuristic | Zajonc (1980, 1984) Slovic (1987); Kahneman and Ritov (1994) Finucane et al. (2000) | Judgments shaped by positive/negative emotions, influencing risk perception and decisions |
| Mimetic Bias | Tarde (1901); Kirman (1993) | imitating the behavior of others without consideration for one's own information |

**Table 1: Categorization of Cognitive Biases by Kahneman's System 1 and System 2**

| | | |
|---|---|---|
| Noise Bias | Kahneman et al. (2021) | Undesirable random variability in judgments. |
| **System 2 Biases (analytical, slow, rational thinking)** | | |
| Confirmation Bias | Wason (1960);<br>Tversky & Kahneman (1974, 1982) | Seeking or interpreting information to confirm existing beliefs. |

Source: Author

The demonstration of these biases has prompted a reassessment of the optimism with which economists constructed *homo economicus*, and of the narrowness of such a conception. Collectively, these contributions undermined the foundational assumption of all models asserting the benefits of deregulated markets. Everything rested on the utilitarian *homo economicus*, whether in the labour market, the money market, goods and services markets, or financial markets. In all these contexts, price alone — assumed to contain all relevant information — was deemed sufficient to guide behaviour. By showing that the average individual is *not* a *homo economicus*, behavioral finance effectively invalidated the very basis of market efficiency. Any form of efficiency presupposes the ability of a majority of agents to form rational expectations — which, in finance, means focusing on fundamental analysis to determine the "true price." If this condition no longer holds, the resulting price cannot be assumed correct.
This also reveals a deeper conceptual issue: even in agent-based models developed within behavioral finance, the same underlying assumption of standard theory remains — namely, that the aggregation of individual behaviour *constitutes* the crowd, and that the micro-level capacity (or failure) of self-regulation automatically generates macroeconomic stability (or macroeconomic instability).

Without necessarily rejecting this micro-to-macro inference mechanism, other scholars within behavioral finance have adopted a more explicitly macroeconomic approach. They study instead the *aggregate consequences* of the absence of individual self-regulation — which, unsurprisingly, manifests as the absence of macroeconomic self-regulation.

### 1.2. To financial instability as an outcome of aggregated biases

Here, we start from the thesis which, in our view, underlies Robert Shiller's *Irrational Exuberance* (2000), in order to integrate the macroeconomic contributions of behavioral finance. First, it is worth noting that in a 2009 interview following the publication of *Animal Spirits* (Akerlof and Shiller, 2009), the journalist asked Robert Shiller whether the book referred to John Maynard Keynes, "who first wrote about the importance of 'animal spirits' or human psychology in relation to the Great Depression?" Not surprisingly, Shiller responded explicitly that "Yes. I think Keynes was right, although we have a better understanding of it now, 70 years later — because of more research in psychology and other social sciences" (Downing, 2009).
Several passages in *The General Theory* (1936) indeed call for the use of psychological tools to understand economic behaviour and its macroeconomic consequences. This is the case, for example, with the psychology of the entrepreneur who must form expectations concerning the marginal efficiency of capital on the basis of effective demand and the interest rate. It is also the case with Keynes's "fundamental psychological law," which posits a declining marginal propensity to consume with rising income. It similarly applies to the interest rate, a highly conventional variable, and finally to speculation: The speculator is one "who tries to guess better than the crowd how the crowd will behave" (Keynes, 1936, p. 157). Keynes consistently emphasised "the importance of emotions, intuition, feelings, and moods" (Schettkat, 2018, p. 42) in decision-making under uncertainty. Alongside this reminder of the inherently gregarious nature of human beings, Keynes's animal spirits were striking because they highlighted — much like cognitive biases — a fundamental propensity toward action, often preceding deliberate reflection: "Our decisions to do something positive can only be taken as a result of animal spirits – of a spontaneous urge to action rather than inaction" (Keynes, 1936, p. 161). It is therefore justified to regard Keynes as practising what would today be described as behavioral economics—and behavioral finance—and explicitly advocating its use (Schettkat, 2018). It is equally evident that Keynes's famous "beauty contest" provided a conceptual foundation for understanding the psychology of the speculator. In short, "He (Keynes, RS) was a true forerunner of behavioral finance" (Thaler 2015, p. 209, in Schettkat, 2018, p. 9).

With this background in place, we can now turn to the work of Robert Shiller, who was likewise a pioneering figure in behavioral finance but at a more macroeconomic level of analysis. As early as 1980, Shiller, together with Grossman, showed that the volatility of stock price indices was far too large to be explained by any reasonable updating of fundamentals, and that it instead reflected fluctuations in risk aversion (Shiller and Grossman, 1980). His subsequent work would formally demonstrate that prices regularly, significantly, and sometimes persistently deviate from their fundamental value, and would seek to explain these deviations. Moreover, the recurring volatility of asset prices itself indicates that such deviations can no longer be dismissed as mere "anomalies"; they stem fundamentally from the workings of human psychology, which intervenes at the level of the market as a whole and renders it intrinsically unstable.

The endogenous propensity of financial markets toward instability had already been identified by Minsky, who, drawing on Keynes, grounded his analysis in financial history (Minsky, 1977; 1986). Shiller (1980; 1984; 1988) thus provided empirical support for Minsky's *financial instability hypothesis*, which sounded an early warning to central banks about the "paradox of tranquillity." This warning was reiterated shortly thereafter by Charles Kindleberger in *Manias, Panics and Crashes* (1978). In *Irrational Exuberance* (2000), Shiller consolidated decades of research on the psychological and macroeconomic functioning of financial markets and set out a comprehensive account of these mechanisms. Among the structural sources of price instability, he identifies — almost as an implicit homage to Minsky, though without citing him explicitly — the Ponzi process that emerges once a threshold level of confidence has been reached. Whereas Minsky adopts from the outset a macroeconomic perspective by "zooming out," Shiller explains the same phenomena from the bottom up, showing how individual behaviours, once aggregated, generate macroeconomic and macropsychological outcomes.

In *Irrational Exuberance*, Shiller attributes financial instability to structural, cultural, and psychological factors. Among the cultural and psychological determinants, he assigns a central role to storytelling — transmitted through word-of-mouth communication — together with the human need for ex post justification, and he analyses the role of mass media in creating and disseminating the narratives that eventually become conventional opinion.

In one sentence — and this is, in our view, the most synthetic formulation of his thesis — Shiller shows that: 1) the herding behaviour arise from information cascade, 2) that arise from word of mouth, 3) which arise from medias, 4) because of their stories telling. This constitutes the analytical structure of our non-exhaustive literature review.

***Herding***

Herding behaviour naturally echoes the animal spirits and the beauty contest mechanism described by Keynes (Devenow and Welch, 1996), reflecting the essentially gregarious nature of human beings, and materialising in the tendency to do "what others are doing rather than using their information" (Banerjee, 1992, p. 1). It can be observed in all living beings and, unsurprisingly, in financial markets as well. In financial settings, herding manifests when everyone sells (or buys) at the same time because everyone expects everyone else to sell (or buy). This is precisely how a convention (either bullish or bearish) takes shape (Orléan, op. cit.).

At the individual level, such behaviour can in fact be rational under uncertainty, as Keynes had already shown — for instance, when individuals believe that others possess relevant information that they themselves do not (Banerjee, 1992). At the macroeconomic level, this synchronisation of behaviours explains the irrational magnitude of price movements relative to what would be acceptable under the EMH (Bikhchandani and Sharma, 2001). Crucially, herding is not limited to noise traders — often depicted as naïve or irrational amateurs (Fama, 1970; Friedman, 1953; Shleiffer and Summers, 1990) — but also characterises institutional investors, who are supposedly professional and therefore rational (Lakonishok, Shleifer and Vishny, 1992).

***Attention cascade and Informational cascades***

The "attention cascade" (Shiller, 2000, p. 79) precedes the informational cascade: attention becomes expectant and sharply focused. As Shiller explains (op. cit., p. 148), "a fundamental observation about human society is that people who communicate regularly with one another think similarly" and "if the millions of people who invest were all truly independent of each other, any faulty thinking would tend to average out, and such thinking would have no effect on prices. But if less-than-mechanistic or irrational thinking is in fact similar over large numbers of people, then such thinking can indeed be the source of stock market booms and busts." He further notes that "the social attention mechanism generates a sudden focus of the attention of the entire community on matters that appear to be emergencies. Thus, to return to the epidemic model, the infection rate may suddenly and dramatically increase. A sudden major move in the stock market is one of those events that pushes aside all other conversation" (op. cit., p. 165).

This phenomenon then gives rise to informational cascades, whose effects on asset prices have been empirically documented (Scharfstein and Stein, 1990; Welch, 2000; Bikhchandani, Hirshleifer and Welch, 1992; Cont and Bouchaud, 2000). An informational cascade functions like a large-scale informational epidemic, steering herd-like behaviour by anchoring expectations. In the case of herding, individuals do not necessarily ignore the information they possess; in an informational cascade, by contrast, individuals *deliberately* disregard their own information (Smith and Sørensen, 2000, in Çelen and Kariv, 2003). In other words, and at the risk of simplifying somewhat, the informational cascade can be understood as a higher-order form of herding — a broader phenomenon which may have multiple causes. The imagery shifts, therefore, from a simple "cascade" to a full-fledged informational avalanche, triggering a behavioural avalanche that sweeps away any deviant behaviour. Such cascades are as short-lived as a fashion and as exponential as an epidemic. They do not depend on a flood of new information: as Bikhchandani, Hirshleifer and Welch note, it may be nothing more than "little new information (…) underlying circumstances have changed (whether or not they really have) (…) and lead to rapid and short-lived fluctuations such as fads, fashions, booms and crashes" (op. cit., p. 2). It is precisely this "little" piece of information that captures collective attention and triggers mass rushes, upward or downward, thereby contributing to the emergence of fat tails in asset returns (Cont and Bouchaud, op. cit.).

***Emotional contagion***

A brief parenthesis is necessary here in relation to the development of Shiller's argument, to emphasise that informational cascades are rooted in the human tendency toward emotional contagion. As Hess and Blairy note, "imitation is a cognitive process, whereas mimicry and emotional contagion operate at an unconscious level. (…) Mimicry constitutes an expressive response, while emotional contagion refers to an affective state." (2001, *in* Van Hoorebeke, 2007, p. 11). This phenomenon was extensively studied by Hatfield and her colleagues. In its primitive sense, emotional contagion is defined as "the tendency to automatically mimic and synchronize facial expressions, vocalizations, postures, and movements with those of another person's and, consequently, to converge emotionally" (Hatfield, Cacioppo and Rapson, 1992, pp. 153–154). More broadly, "emotional contagion refers to the process in which an observed behavioral change in one individual leads to the reflexive production of the same behaviour by other individuals in close proximity, with the likely outcome of converging emotionally" (Panksepp and Lahvis, 2011).
Once again, emotion precedes and drives action. Johnson and Tversky (1983) were the first to show the extent to which mood affects risk assessment. At a more macroeconomic level, DeLong et al. (1990) demonstrated the role of collective emotions and sentiments in financial decision-making. Au et al. (2003, p. 1) found that "a good mood" leads to less precise decisions than a "bad mood". The "socioeconomic hypothesis" holds that emotions spread through interpersonal interaction and ultimately determine behaviour (Prechter, 1999). Lo and Repin (2002) observed that market volatility affects cardiovascular variables, indicating that emotional contagion in financial markets may be transmitted through prices — or rather through the observation of prices volatility. Ultimately, collective mood manifests in collective behaviour (Prechter, 2003). Nofsinger (2005, p. 157) argues that "the stock market is a measure of social mood," and not the other way around. Prices in financial markets reflect collective mood because they track sentiment; they do not lead it. Moreover, the volatility of prices is proportional to the extremity of social mood, such that the more intense the mood, the more extreme the market fluctuations. Given the high degree of interconnection across markets, such contagion can easily become international and be transmitted through capital flows, even though "there are surely others [mechanisms], including word-of-mouth and the media" (Baker, Wurgler and Yuan, 2012, p. 3).

***Communication: Vector of Emotional Contagion***

Emotional contagion, much like an epidemic, spreads through communication. The role of communication as a source of psychological transmission falls within the scope of what Shiller calls "narrative economics," which he defines as "the study of the spread and dynamics of popular narratives, the stories, particularly those of human interest and emotion, and how these change through time, to understand economic fluctuations" (Shiller, 2017, p. 967). Our interest here lies in the mechanisms through which these narratives propagate, and, following Shiller, we interpret them as epidemics that spread rapidly and with high emotional intensity.

*Word-of-mouth*

Word-of-mouth is a major trigger of emotional contagion and herding: "The conventional media—print media, television, and radio—have a profound capability for spreading ideas, but their ability to generate active behaviors is still limited. Interpersonal and inter-active communications, particularly face-to-face or word-of-mouth communications, still have the most powerful impact on our behavior" (Shiller, 2000, p. 154). Word-of-mouth communication is indeed epidemic in nature. Shiller draws upon mathematical models of epidemic dynamics, which show propagation accelerating exponentially once a threshold is crossed, and compares this to the viral spread of information documented by sociological research.
However, the problem with word-of-mouth is that the transmission of a message from person to person eventually affects its content, or even its nature, distorting it in the process (Allport and Postman, 1947, *in* Shiller, op. cit.). This phenomenon can amplify the perceived importance of the message. Shiller had already shown that interpersonal communication plays a decisive role in shaping behaviour (Shiller and Pound, 1989). But at that time, technological means of remote communication were still limited. A few years later, Shiller (2000) confirmed the significant role of new technologies in facilitating emotional transmission — including email, telephone, and, even more today, cell phones and videoconferencing, which accelerate direct interpersonal communication.

*The press*

In addition, the media constitute a major source of emotional generation and emotional contagion. Although Shiller qualifies the role of the media relative to word-of-mouth communication, he nonetheless observes that “significant market events generally occur only if there is similar thinking among large groups of people, and the news media are essential vehicles for the spread of ideas. (…) However, a careful analysis reveals that the news media do play an important role both in setting the stage for market moves and in instigating the moves themselves” (Shiller, op. cit., p. 71). Quinn and Turner (2020) similarly emphasise the systematic role of the media and financial elites in the emergence of asset price bubbles.

With respect to the press, Kindleberger (1978) extensively highlighted its function in amplifying and spreading speculative euphoria. De Bondt and Thaler (1985) also underscored the role of “news” in both underpricing and overpricing of financial assets. Cutler, Poterba and Summers (1988) did not find a statistically significant causal relationship between macroeconomic news and return variance, because only one third of return variance can be explained by macroeconomic news. However, some years later, Tetlock (2005), using textual analysis of the Wall Street Journal, showed that “high values of media pessimism induce downward pressure on market prices; and unusually high or low values of pessimism lead to temporarily high market trading volume” (Tetlock, 2005, p. 30). More importantly, Tetlock stresses that this pessimism is not reflected in fundamentals and does not originate from them.

*Forums*

Regarding the role of online discussion forums, Antweiler and Frank (2004, p. 1) analysed “more than 1.5 million messages posted on Yahoo! Finance and Raging Bull about the 45 companies in the Dow Jones Industrial Average and the Dow Jones Internet Index.” They found only a weak relationship between optimistic messages and returns, but they did identify a clear link between the volume of messages and trading volume.

*Social networks*

More broadly, a number of studies suggest that social networks play a significant role in emotional contagion, even in the absence of direct interpersonal interaction. By experimentally manipulating Facebook news feeds, Kramer, Guillory and Hancock (2014, p. 1) showed “that emotions expressed by others on Facebook influence our own emotions, constituting experimental evidence for massive-scale contagion via social networks. This work also suggests that, in contrast to prevailing assumptions, in-person interaction and nonverbal cues are not strictly necessary for emotional contagion.”

Ferrara and Yang (2015) confirmed these findings by examining the impact of positive and negative Twitter content on users’ emotional states, as reflected in the tone of their subsequent posts. They further observed that positive content has a stronger effect than negative content, because it also influences individuals who are ordinarily less susceptible to emotional contagion. This result has been fully corroborated by Garcia (2024), who shows that sentiment transmitted through Twitter influences the herding behaviour of contributors on Estimize.

***Stories telling***

What, then, do most of these highly influential communications actually contain? Hard evidence? Objective information? No — merely stories, much like legends or tales that alternately inspire hope or instil fear. Shiller (2000, p. 139) reports the experiment by Pennington and Hastie (1992), who "have shown the importance of stories in decision making by studying how jurors reached decisions in difficult cases." This observation had already been made by MacKay (1852), who noted that the collective delusions he studied were fuelled by the circulation of narratives and illustrative examples. Shiller (op. cit., p. 141) also refers to the work of Shafir, Simonson and Tversky (2000) concerning the ex post rationalisation of decisions, after two groups chose — for opposite reasons — the more "enriched" description over the less detailed one. In practice, we often need to construct a coherent story to justify our behaviours and decisions, and the more elaborately detailed this story is, the more we tend to believe that it contains supporting evidence for our choice. With the example of the bestseller by Thomas Stanley and William Danko, *The Millionaire Next Door: The Surprising Secrets of America's Wealthy* (1996), Shiller shows that this highly influential book encouraging individuals to invest their savings contains almost no technical detail, demonstrative evidence, or concrete investment advice. It succeeds because it tells a success story rooted in imagery like the American dream and the Self-made man. It activates imagination and hope, so that the reader feels capable of flight long before any wings have been proven to exist. As Shiller concludes, "Thus the likelihood of any event affecting market prices is enhanced if there is a good, vivid, tellable story about the event" (op. cit., p. 161). The success of such works illustrates once again that the absence of face-to-face communication does not in any way prevent emotional contagion.

Thus, the core mechanism underlying the phenomena observed during boom-and-bust episodes is interaction — even when indirect or mediated at a distance. The trader, regardless of institutional status, may sit alone behind one or several computer screens, but he is not *socially* isolated, and taking isolated excellence as a normative benchmark is misleading. As Kirman notes, "the behavior of the group as a whole cannot be inferred from analyzing one of the identical individuals in isolation. Without taking explicit account of the interaction between individuals, the group behavior observed during the experiments cannot be explained. This observation is of interest to economists, since similar phenomena have been observed in markets" (Kirman, 1993, p. 2, about ants herding).

Finally, "Samuelson's [1998] dictum that the stock market is micro efficient, but macro inefficient" (*in* Nofsinger, 2005, p. 9) appears only partially correct. On the one hand, it is not strictly accurate, since behavioral finance in its microeconomic dimension has already shown that individual rationality is impaired by cognitive biases — thereby challenging micro efficiency itself. On the other hand, the dictum is meaningful insofar as it acknowledges the gap between the microeconomic level, associated with individual rationality, and the macroeconomic level, associated with collective rationality. The crowd has its own logic, which individual logic does not necessarily capture.

The macroeconomic approach in behavioral finance tends to imply that the aggregation of bounded rationalities leads to macro inefficiency. While this is intuitively true — since the sum of individual inefficiencies should indeed produce collective inefficiency — our intention here is not to refute this postulate but to extend it. Our aim is to propose a complementary macroeconomic explanation of inefficiency, as an additional layer of understanding, in the hope of modestly enriching the field.

**Table 2: Microeconomics and macroeconomics behavioral approaches**

| Approach | First author(s) | Key contribution |
|---|---|---|
| Micro (individual, cognitive biases) | Simon (1955, 1957) | Bounded rationality; agents rely on satisficing. |
| | Kahneman & Tversky (1974, 1979) | Heuristics, biases ; Prospect Theory with loss aversion, reference dependence. |
| | Thaler (1985) | Mental accounting, endowment effect, myopic loss aversion —> Explains anomalies. |
| | Shefrin & Statman (1984, 2000) | Disposition effect: Sell winners too early, hold losers too long; behavioral portfolio theory. |
| | Barberis, Shleifer & Vishny (1997) | Conservatism (underreaction) and representativeness (overreaction) drive return patterns. |
| | Daniel, Hirshleifer & Subrahmanyam (1998) | Overconfidence, self-attribution → Excess volatility, momentum–reversal. |
| Macro (crowd, herding, contagion) | Shiller (1981, 2000) | Word-of-mouth, story telling and emotional contagion |
| | Keynes (1936); Akerlof & Shiller (2009) | Animal spirits |
| | Banerjee (1992); Kirman (1993) | Herding |
| | Bikhchandani, Hirshleifer & Welch (1992) | Informational cascades drive herding |
| | Hatfield (1992, 1994) | Emotional contagion |
| | Shiller (2000) | Impacts of direct communication on collective emotion: Word-of-mouth |
| | De Bondt & Thaler (1985); Antweiler & Frank (2004); Tetlock (2005); Kramer, Guillory & Hancock (2014); Ferrara & Yang (2015) | Impacts of indirect communication on collective emotion: News, press and social medias |

Source: Author

In the following section, we shift the focus explicitly to boom-and-bust periods and bracket out all considerations related to the individual — whether rational or cognitively biased — in order to examine instead the individual *as absorbed into the crowd*. Under these exceptional circumstances, which justify this analytical framing, the crowd acquires a “mental unity” of its own (Le Bon, op. cit.). This zooming-out reveals that the crowd is no longer the average or the sum of the individuals who compose it.

We will show that the defining characteristics of financial markets identified by the macroeconomic strand of behavioral finance are corroborated through a distinct analytical lens — that of crowd psychology.

Ultimately, what we develop here reinforces Shiller’s claim that “mass psychology may well be the dominant cause of movements in the price of the aggregate stock market” (Shiller, 1984, p. 457).

## 2. Financial markets during boom and bust: a paradigmatic Le Bon-style crowd

This second part begins by outlining several links established in the literature between financial markets under bubble-and-crash conditions and what is commonly described as crowd phenomena. Many scholars have, in effect, implicitly treated markets as a crowd, even if they have not — to our knowledge — fully developed the analysis in those terms. Our objective here is to formalise this connection by grounding it in Gustave Le Bon’s notion of the “psychological crowd”, whose defining characteristics align closely with the macroeconomic mechanisms identified in the first part of this article. This correspondence, strengthened through a focus on the macroeconomic dimension, provides support for the relevance of such a framework.

### 2.1. The crowd-like nature of financial markets in economic thought

A number of authors who have studied the functioning and psychology of financial markets during boom-and-bust episodes have explicitly described them as a crowd. The term has often appeared in descriptive analyses, but rarely within an explicit theoretical framework.

Charles MacKay was among the first to do so, using the term directly in the title of his 1852 classic *Memoirs of Extraordinary Popular Delusions and the Madness of Crowds*. In it, he describes how human beings are prone to irrational behaviour and can be easily swept up in collective illusions and manias when absorbed into crowd phenomena, as if they became captive to the crowd itself. Drawing on the example of the Mississippi Company bubble in France, he illustrates how collective illusions can generate widespread folly and lead to severe economic crises, as individuals are carried away by an illusion of omnipotence and neglect fundamental risks.

Of course, Keynes (1936) repeatedly referred to mass psychology throughout *The General Theory* to characterise macroeconomic and financial phenomena. It is worth noting that Keynes was familiar with Freud’s work on analytical psychology and crowd psychology (1921), and drew inspiration from it in his analysis of macroeconomic and financial behaviour (Dostaler and Maris, 2009). In financial markets, Keynes explains for example that “he who attempts it [talking about investment strategy, compared to speculative strategy] must surely lead much more laborious days and run greater risks than he who tries to guess better than the crowd how the crowd will behave…” (p. 157). He adds: “They are concerned, not with what an investment is really worth to a man who buys it ‘for keeps’, but with what the market will value it at, under the influence of mass psychology, three months or a year hence. (…) Thus the professional investor is forced to concern himself with the anticipation of impending changes, in the news or in the atmosphere, of the kind by which experience shows that the mass psychology of the market is most influenced" (p. 99). And further, "This is closely analogous to what we have already discussed at some length in connection with the marginal efficiency of capital. Just as we found that the marginal efficiency of capital is fixed, not by the 'best' opinion, but by the market valuation as determined by mass psychology, so also expectations as to the future of the rate of interest as fixed by mass psychology have their reactions on liquidity-preference" (p. 108). Finally, regarding asset pricing: "a conventional valuation which is established as the outcome of the mass psychology of a large number of ignorant individuals is liable to change violently as the result of a sudden fluctuation of opinion due to factors which do not really make much difference to the prospective yield; since there will be no strong roots of conviction to hold it steady." At first glance, this may appear surprising since Keynes, as a financial market participant, was broadly a fundamentalist investor — seeking to anticipate, better than the crowd, future macroeconomic fundamentals (Accominotti and Chambers, 2017). In reality, however, Keynes was attempting to anticipate which fundamentals would capture the crowd's attention — the famous *attention cascade* — and thereby trigger emotional contagion.

Kindleberger (1978) can be said to "psychologise" Minsky's (1986) paradox. Minsky himself uses the term *crowd* only rarely. One of the few places where it does appear is in his discussion of bank runs, which he sees as the clearest signal that a crisis is underway: "in a modern banking environment, bank runs are much more polite affairs than in earlier times, when clamoring crowds would gather outside a bank in distress to exchange deposits for currency" (Minsky, 1986, p. 63). Yet what he describes in contemporary financial markets is structurally the same phenomenon. Even when the term is absent, one can feel that when he writes about an "unstable system," or "financial behavior" (p. 220), or asks why "the behavior of the economy changes" (p. 219), he is implicitly referring to a crowd phenomenon — sudden, amplified, and collective by nature. When he expresses concern about "the cumulative decline," or about a lender of last resort who "stand[s] aside and allow[s] market forces to operate" (p. 44), the underlying dynamic resembles that of a crowd left to itself, deprived of leadership and therefore vulnerable to panic.
Kindleberger, by contrast, states this dynamic explicitly: "in every mania, some get rich and attract a following. The crowd imitates, and the mass movement reinforces itself" (p. 40). Later in the cycle, "mass psychology takes over in the later stages of the mania, as each participant thinks only of joining the crowd before it is too late" (p. 89). Finally, when the reversal sets in, "the critical stage is reached when the rush to get out turns into a panic… the crowd stampedes" (p. 96).

Shiller corroborated all these views as early as 1984. In *Irrational Exuberance* (2000, p. 152 and 155), he reiterates this view, noting that "the volatility of the market is amplified by crowd reactions, as people respond to the same stories and the same signals," and that "speculative bubbles appear to be caused by a feedback loop from price increases to increased investor enthusiasm, involving contagion effects as in an epidemic, and amplified by the news media in a way that has parallels with mass delusions."

The French "école des conventions," *Convention theory*) represented most notably by the work of André Orléan, draws on sociology and on the Keynesian understanding of financial markets to arrive at conclusions similar to those of behavioral finance. As Orléan (2001, p. 105) puts it, « loin d'être un ensemble d'individus séparés et indépendants, le marché ressemble plus à une communauté fortement interconnectée, voire même, lors de certains épisodes spéculatifs, à une foule abandonnée à son propre mouvement. La contagion des comportements y joue un rôle central ».
Thomas Lux (1995) formally modelled herd behavior in speculative markets. He argues that "it is postulated furthermore that the speculators' readiness to follow the crowd depends on one basic economic variable" (pp. 881–882).

In a multi-agent based model, Bak, Paczuski and Shubik (1996, p. 1) confirm Shiller's thesis by showing that mutual imitation generates "a crowd effect" which in turn produces large price fluctuations.
Bouchaud and Cont (op. cit., p. 1) likewise conclude that "returns may correspond to collective phenomena such as crowd effects or 'herd' behavior."
Fenzl, Brudermann and Pelzmann (2013, p. 421) analyse "how the mass psychological perspective contributes to the discussion on the origins of the recent global economic and financial crisis." They attribute the emergence of crowd phenomena to long-term structural and social changes, which intensified such behaviours, and to the abrupt unwinding of accumulated debt, which triggered a panic dynamic.
Hansen (2017) devoted his PhD, *Crowds and Speculation: A Study of Crowd Phenomena in the U.S. Financial Markets 1890 to 1940*, to establishing a historical and empirical link between boom-and-bust episodes and crowd psychology, particularly as conceptualised by Le Bon (1895) and Tarde (1901).
Almansour (2023, p. 4) treats herding as a cognitive bias in its own right, but nonetheless confirms that "many investors tend to follow the crowd or exhibit overconfidence biases when making investment decisions."
Xu (2025, p. 1) similarly echoes the views of Le Bon (1895), Tarde (1901) and Freud (1921) in noting that "the notion of 'wisdom of crowds' postulates that collective decision-making often outperforms individual judgements. However, in financial markets, this collective intelligence can falter, leading to inefficiencies and anomalies," although in his analysis these inefficiencies stem primarily from informational mechanisms.

For these three pioneers of crowd psychology — Le Bon (1895), Tarde (1901) and Freud (1921) — a crowd can indeed be more productive or effective than the individual, but only when it is structured and organised around a leader. Otherwise, "the intellectual abilities of men (...) fade away. Heterogeneity is drowned out by homogeneity, and unconscious qualities dominate. This pooling of ordinary qualities explains why crowds are incapable of performing acts that require high intelligence" (Le Bon, op. cit., p. 19).

### 2.2. Financial markets as a psychological crowd under particular circumstances: Boom-and-bust

Modelling the functioning and characteristics of financial market overreactions while treating the market as an autonomous entity, distinct from the individuals who compose it, is in principle impossible. Doing so would amount to modelling an agent that is irrational, unconscious, suggestible, highly sensitive, and prone to sudden impulsive shifts. Such a model can therefore only be qualitative in nature and supported by empirical observation.

#### 2.2.1. The Le Bon's psychological crowd

Within social psychology, the distinctive feature of crowd psychology is that it emphasizes (1) that the crowd constitutes a single unit, driven by "herd instinct, group mind – which is not displayed in other situations" (Freud, 1921, p. 50), and (2) the central role of the leader, since in psychological terms, there can be no crowd without a leader. This is also why Freud preferred the concept of *mass* to Le Bon's term *foule*. The notion of *mass* conveys the idea of a unified block that suppresses the heterogeneity of the individuals who compose it. In what follows, we will therefore use the terms *mass* and *crowd* interchangeably.

But what, more precisely, is a crowd? Le Bon answers this question as follows: "From a psychological point of view, (…) under certain given circumstances, and only under those circumstances, an aggregation of individuals acquires new characteristics that are profoundly different from those of the individuals who compose it. The conscious personality fades away, and the feelings and ideas of all the members become oriented in the same direction. A collective soul is then formed — transitory, no doubt, but presenting very distinct features. The group thus becomes what, for lack of a better term, I shall call an organized crowd, or, if one prefers, a psychological crowd. It becomes a single being and is subject to the psychological law of the mental unity of crowds." (Le Bon, op.cit., p. 16).
Le Bon further explains that he is primarily concerned with "heterogeneous crowds," which may be anonymous, such as street crowds, or non-anonymous, such as juries or parliamentary assemblies. He distinguishes these heterogeneous crowds from "homogeneous crowds," such as sects, castes, or social classes (p. 83). Freud (1921) draws a parallel distinction between natural crowds, in which no clearly identifiable leader exists, and artificial crowds, which are organised around a charismatic leader.
In the context of boom-and-bust episodes, which we regard as precisely the kind of exceptional circumstances described by Le Bon, financial markets constitute a heterogeneous and natural psychological crowd.

We might alternatively have chosen, as a point of reference for comparison with financial markets, Tarde's (1901, p. 9) concept of the public, defined as "a purely spiritual community, like a scattering of physically separated individuals whose cohesion is entirely mental." However, even for Le Bon, "thousands of separate individuals can, at a given moment, under the influence of certain violent emotions, such as a major national event, acquire the characteristics of a psychological crowd" (Le Bon, op. cit., p. 17). Thus, physical proximity and direct communication are not necessary conditions for the formation of a crowd in Le Bon's theory either. This has been further confirmed by Carsten Stage, who identifies "online crowd (…) characterized by intense affective unification. (…) In other words, 'online crowding' refers to the affective unification and relative synchronization of a public in relation to a specific online site" (Stage, 2013, p. 6).

Secondly, financial markets correspond more closely to Le Bon's conception of the crowd than to Tarde's notion of the public. Tarde's public is a more cognitively elevated form of collectivity — a kind of higher-order, more "sober" crowd that is not swayed by just any external shock. On financial markets, by contrast, shocks of all origins routinely generate overreactions.

Le Bon identifies two "special characteristics" of the crowd.
First, the crowd is characterised by the emergence of a "collective soul." As Le Bon explains (op. cit., p. 18): "The most striking fact presented by a psychological crowd is this: whoever the individuals that compose it may be, however similar or dissimilar their lives, occupations, character, or intelligence, the mere fact that they have been transformed into a crowd endows them with a sort of collective soul. The psychological crowd is a provisional being, composed for a moment of heterogeneous elements that have been fused together, just as the cells of a living body combine to form a new organism exhibiting characteristics quite different from those possessed by each cell taken separately." This collective soul manifests itself, as we shall see, through the erasure of individual personalities.
Second, the crowd is governed, not by consciousness, but by the unconscious: "Now, these general traits of character, governed by the unconscious and shared to nearly the same degree by most normal individuals of a given group, are precisely those that become pooled together within a crowd. In the collective soul, the intellectual faculties of individuals — and thus their individuality — fade away. The heterogeneous is submerged into the homogeneous, and unconscious qualities prevail" (op. cit., pp. 18–19). In other words, the crowd is, for Le Bon, the social form assumed by the unconscious — the repository of humanity's primitive roots. For Freud, it is likewise the place where repressed drives re-emerge. And for Kahneman, it is the domain of System 1: rapid, automatic, and largely uncontrollable. The terminology differs, but the consequence is identical: the crowd is, before anything else, instinctive.

Moreover, the distinctive feature of crowd psychology is captured in Le Bon's well-known formulation: "Thus, the disappearance of the conscious personality, the predominance of the unconscious personality, the turning of ideas into acts by means of suggestion and contagion, and the tendency to immediately transform suggested ideas into action, such are the principal characteristics of the individual forming part of a crowd." (Le Bon, op. cit., p. 20).
For Reicher (2004, pp. 4–5), the transformation of the individual within a crowd has a partly deliberate component, insofar as individuals "shift from personal identity (what makes me as an individual distinctive from other individuals) to social identity (what makes my group distinctive compared to other groups)." Yet in both conceptions, a transformation undeniably occurs. I retain Le Bon's notion of a temporary disappearance of individual identity, given the uncertainty and difficulty with which individuals involved in crowd phenomena later account for their own behaviour (Shiller, 2000).

### 2.2.2. Financial markets as Le Bonian crowds: structural similarities and identical outcomes

We now return to Le Bon's key characteristics of the crowd and examine them one by one, in order to explain and corroborate — from a different analytical perspective — the features of financial markets that have been extensively documented by macroeconomic approaches within behavioral finance during episodes of manias and crashes. For a brief comparative synthesis, see Table 3, p. 29.

***The disappearance of conscious personality and the predominance of the unconscious***

Gustave Le Bon explains that once immersed in a psychological crowd, the individual undergoes a transformation. The conscious part of the self — that is, one's rational faculty — evaporates into the "effluves" of the crowd (op. cit., p. 20), allowing the unconscious part — instinctive, impulsive, and emotional — to take control of both anticipation and action.
This corresponds precisely to what Stanovich and West (2000), and later Kahneman (2002), conceptualised as System 1: a mode of cognition that is rapid, intuitive, above all unconscious, and therefore not subject to reflective control. If the crowd appears irrational, it is because it *does not reason*; once the relevant conditions are in place, it reacts automatically to stimuli — what Le Bon calls *suggestions*. On financial markets, such conditions are those characterised by urgency, or what Lo and Repin (2002) describe as *transient market events*: relatively sudden occurrences that compel immediate action, leaving no time for deliberation, and marked by pressure and emotional arousal. This applies equally to speculative bubbles and to crises. System 1 is therefore the unconscious dimension of cognition that Kahneman (2012, p. 87) calls "the stranger within us". Those familiar with Freud may recognise here *the unconscious* in the Freudian sense: the impulsive "stranger" that governs much of our behaviour. This unconscious component, activated through immersion in crowd phenomena, has been extensively documented in the psychological and especially the cognitive literature, which has "confirmed the Freudian insight about the role of symbols and metaphors in unconscious associations" (op. cit., p. 90). For example, Kahneman recounts an experiment in which one group of subjects read a set of words associated with old age (although the word *old* was never explicitly mentioned), while a control group read unrelated words. The real object of observation came afterward, when participants were asked to walk to another room: those primed with words related to old age walked more slowly, exactly as the experimenters had predicted. As we have seen in Part I, nearly all of the cognitive biases documented in economics and psychology belong to System 1. The crowd, conceived as an entity distinct from the individuals who compose it, is therefore itself subject to these same unconscious heuristic processes.

In facts, the exceptional circumstances in which the psychological crowd emerges give rise to the phenomenon of *de-individuation*, which refers to the loss of individuality when the self dissolves into the collective "effluves" of the mass. In psychology, Festinger, Pepitone and Newcomb (1952, p. 1), pioneers on the topic, describe it as follows: "a group phenomenon which we have called de-individuation has been described and defined as a state of affairs in a group where members do not pay attention to other individuals qua individuals, and, correspondingly, the members do not feel they are being singled out by others. The theory was advanced that this results in a reduction of inner restraints in the members and that, consequently, the members will be more free to indulge in behavior from which they are usually restrained."
This de-individuation effect has been observed in a number of small-group experiments. Although severely and legitimately criticised for methodological fraud and for its psychological consequences (Le Texier, 2019), the Stanford prison experiment (Haney and Zimbardo, 1971), which placed two groups of individuals in a simulated prison setting, illustrated how, once immersed in their respective groups, participants "became" their assigned roles, appearing "dissolved" into them — either as irrationally submissive prisoners or as irrationally sadistic guards. The protocolic manipulation by Zimbardo might suggest that the experiment should be reclassified primarily as a study in obedience to authority; nevertheless, among the guards especially, the group dynamic clearly played a decisive role.
In sociology, the prevailing term for this phenomenon is *depersonalization*. As Lindholm (1992, p. 14 and 16) notes, "Durkheim, unlike Weber, draws a radical distinction between the goals and character of the group and the goals and characters of the individuals within the group, arguing that 'social psychology has its own laws that are not those of individual psychology' (1966 Suicide: 312) (…) Durkheim thought that an extraordinary altered state of consciousness among individuals in a group, which he called 'collective effervescence' would occur spontaneously (Durkheim 1965: 240-1). This experience is one of depersonalization, and of a transcendent sense of participation in something larger and more powerful than themselves".

This de-individuation bears some resemblance to the findings of Asch's (1951) experiment on conformity, in which, on average, more than one third of participants (37%) gave an incorrect answer to a very simple question when exposed to the influence of a unanimous confederate group, even though all of them (100 percent) had given the correct answer when alone. The task itself draws on System 1 because of its simplicity, while the contradiction with the group's incorrect answer would normally activate System 2. The outcome — a strikingly high rate of submission to the group — shows that belonging, or the *desire* for belonging (driven by fear of rejection), overrides individual rational judgment. In neurobiology, Berns et al. (2005) provided supporting evidence using functional magnetic resonance imaging in situations of peer pressure and exposure to incorrect information, demonstrating that conformity alters neural activity in specific brain regions, thereby offering physical proof of the psychological transformation that occurs in group contexts. De-individuation is thus the cause of the conformity observed by Asch, and becomes far more acute under conditions of pressure — a state commonly referred to as a crowd phenomenon. In a different register, Darley and Latané (1968) showed that the greater the number of people present in a group, the greater the inertia, as each individual expects someone else to take action first. Through this dissolution of individuality into the crowd, the individual *relies* on the crowd.

In the specific circumstances of bubbles and crashes, there is no deliberate strategy at play: the individual does not *choose* to merge into the crowd, nor consciously decide to conform to it. Under conditions of urgency, System 1 — the unconscious, or instinct — takes control and compels alignment with the crowd. This is precisely what emerges from Kindleberger's historical investigations, which repeatedly describe two extreme states: collective hysteria, followed by a wave of general panic. It is also what is captured in Minsky's vocabulary of euphoria and distress within his "paradox of tranquillity." The same logic is reflected in post-crash survey data, in which market participants report that they had no choice but to buy during the boom and sell during the crash — regardless of fundamentals, and regardless of macroeconomic consequences. Shiller (1988) clearly documented this emotional contamination during the dramatic 1987 crash, which bore no relation to the fundamentals prevailing at the time. His survey shows that "20.3% of the individual investors in INDIV and 43.1% of the institutional investors" experienced "difficulty concentrating, sweaty palms, tightness in chest, irritability, or rapid pulse" and concludes that "It is remarkable that such a proportion of the general population reported such specific symptoms of real anxiety at one time." (…) "Among individual investors who sold on October 19, 53.9% reported experiencing contagion of fear" (Shiller, 1988, pp. 11–12). Here, crowd fear quite literally took possession of individuals — remotely — via falling price curves. Shiller (2000) also highlights the confusion and incoherence of ex post explanations, as investors attempt — unsuccessfully — to rationalise behaviour after the fact, as if waking from the hypnotic state described by Le Bon when the individual is absorbed into the crowd.

***Orientation through suggestion***

The unconscious, far more than the conscious mind, is particularly susceptible to suggestion, in the sense that it reacts intensely to it. As Le Bon explains (op. cit., pp. 19–20): "To understand this phenomenon, we must bear in mind certain recent discoveries of physiology. We know now that an individual may be brought into such a condition that, having lost his conscious personality, he obeys all suggestions of the operator who has deprived him of it, and commits acts in utter contradiction with his character and habits. Now, attentive observation seems to show that an individual, when he has for some time been in the midst of an active crowd, is soon placed, by reason of the magnetism emanating from it, or from some other cause of which we are as yet ignorant, in a special state which closely resembles the state of fascination of the hypnotised subject in the hands of his hypnotiser." Freud (1921) further notes that, contrary to popular belief, hypnosis at the time was not only an individual practice but was often carried out in groups. Each subject about to be hypnotised was first exposed to the testimony of others who had already been successfully hypnotised. Hypnosis remains difficult to fully explain, even today, but in essence it consists in implanting an idea into the unconscious through suggestion — an idea that later persists in the background, becoming quasi-obsessional, and which the hypnotised person may eventually act upon long after the hypnotic session itself, in a kind of delayed and programmed enactment.

This process of suggestion is generally carried out by a figure of authority and/or trust (there must always be a link, however tenuous, of an affective nature), and it is never articulated through logic or demonstrative reasoning. On the contrary, it is conveyed through unprepared language: extremely simple, concise, and emotionally charged, often consisting of a single powerful word, striking formula, or evocative image. Only such simple but symbolically strong words — capable of triggering vivid mental imagery — give rise to suggestion. Again, this is consistent with findings in cognitive psychology reported by Kahneman (2012), who concludes that Freudian insights have been empirically confirmed "about the role of symbols and metaphors in unconscious associations" (op. cit., p. 90).

As Le Bon notes, "however neutral we may suppose it to be, a crowd is always in a state of expectant attention, which renders it extremely susceptible to suggestion. The first suggestion expressed implants itself immediately by a process of contagion in the minds of all, and the direction is at once determined."(op. cit., p. 24). He continues: "Whatever the ideas that are suggested to crowds, they can only become dominant on the condition of assuming a very simple form, and presenting themselves under the guise of striking images. There is no connection of logical sequence or analogy between these ideas-images; they may replace one another just as the slides of a magic lantern are withdrawn from the box in which they are piled. One may therefore see in crowds the most contradictory ideas succeed each other" (op. cit., p. 35). This is what Moscovici (op. cit., p. 121) calls "image-based thinking", which he distinguishes from the "critical thinking" of the individual. The crowd's mode of cognition is automatic: it unfolds through stereotyped associations, which are converted into images and then into image-based thinking, before eventually materialising into action (cf. infra).

“How is the imagination of crowds to be impressed? The things that impress it most strongly are those which present themselves in the form of striking and vivid images, freed from all accessory explanation, or accompanied only by a few marvellous facts — a great victory, a great miracle, a great crime, a great hope” (Le Bon, op. cit., p. 37). This is why, for Le Bon — as for Freud, who further developed the central role of the leader — the latter must be charismatic and must never rely on logical argumentation if he wishes to sway the crowd through striking impression rather than reasoning. As Le Bon observes: “The weakness of certain speeches which have had an enormous influence on those who heard them sometimes surprises us when we read them; but we forget that they were intended to convince assemblies, and not to be read by philosophers. The orator in intimate communication with the crowd knows how to evoke the images which enthral it.” (op. cit., p. 37). The term “image” should not be taken exclusively in its literal sense, because “the judicious employment of words and formulas, and their constant repetition, are the means by which crowds are influenced. Used with art, they truly possess the mysterious power formerly attributed to magic spells. They are capable of provoking the most tremendous storms in the souls of crowds, and they can also calm them.” (p. 56). In Le Bon’s view — and this is also the assessment we follow — “logical minds, accustomed to close reasoning, are always surprised at the ineffectiveness of their arguments when they have to do with crowds, and cannot refrain from employing this mode of persuasion; but the failure of their efforts invariably astonishes them.” (op. cit., p. 61).

This is precisely what we recognise in Shiller’s (2017; 2019) *economic narratives*: stories — sometimes frightening, sometimes inspirational — that capture the imagination, thereby triggering the attention cascade and informational cascades. On the manic side, that is, when suggestion gives rise to euphoria, Akerlof and Shiller (2009, p. 182) cite, for example, the advertising campaigns of Merrill Lynch in the late 1990s and early 2000s. Before the 2000 bubble, their communication “one advertisement showed a grandfather and his grandson peacefully fishing together, with the caption: ‘Prepare to be rich. Slowly.’” After the bubble burst, the messaging shifted to “an electronic chip in the shape of a bull, with the caption: ‘Stay connected… and charge!’”(op. cit.). As Thomas Colmard (2013, p. 1) notes in his reading of *Animal Spirits*, these narratives operate as “more general factors operate as collective beliefs, such as monetary illusion or the transmission of stories that are only loosely connected to established facts. (…) Irrationality prevails and is reflected in the profusion of narratives that sustain the myth of an ever-rising housing market”. Such suggestion — simple, even simplistic, and highly concise — echoes the results of Bikhchandani, Hirshleifer and Welch (1992) regarding the ability of a mere *little news* to generate price movements, as well as Shiller (2000) and Tetlock (2005), who show how media sentiment can induce price swings with no connection to fundamentals.

Moreover, suggestion prevents those who have been “hypnotised” in this way from providing coherent ex post explanations of their behaviour. This is also why Shiller (op. cit.) refers to the vague, dreamlike state reported by investors attempting to rationalise after the fact what, in reality, never belonged to the domain of rational deliberation. These narratives, capable of mobilising an expectant crowd, are never rigorously argued, never technical in form: instead, they suggest images — of success, of wealth, of superiority — and can trigger a genuine rush, a collective “gold fever.” The most recent global financial crisis of 2008 has been widely documented, and it clearly displays the “effluves” of crowd hypnosis described by Le Bon: the crowd collectively disregarded the illegality of certain operations, the inevitable insolvency of borrowers, and the magnitude of the catastrophe to come. As Le Bon warns (p. 19): “The individual in a crowd acquires, by the mere fact of its number, a feeling of invincible power which allows him to yield to instincts that, had he been alone, he would perforce have kept under restraint”. Individually, some market participants may well have foreseen the scale of the euphoria and of the subsequent collapse. But once absorbed into the crowd, the overwhelming majority became blinded to the risks: any deployment of System 2 reasoning — which should have produced caution — was short-circuited.

The same process operates under crash conditions. Even if System 2 “knows” that the price has fallen far below what could plausibly correspond to fundamental value, it no longer holds the controls. A single word, a single image, and a wave of panic sweeps through the entire crowd. Within it, each individual is merely a feather caught in the storm. System 1 takes over, and, paradoxically, it is precisely in the moment of panic — when the crowd disintegrates — that the presence of a crowd becomes most visible (Canetti, 1962). At that very moment, each person abruptly rediscovers, not cognitively but instinctively, his or her solitude and vulnerability. In such circumstances, the absence of a leader is fatal, as Freud (1921) observed in descriptions of military collapses following the disappearance of the commanding figure. The suggestion — conveyed here through collapsing prices and pessimistic news — has fixed the direction of movement, but there is no one to reassure and guide the crowd toward a safe exit. Stiglitz (2002) criticised IMF communication during periods of acute financial tension, when a bubble had supposedly reached its peak, but no one knew when it would burst. Radical uncertainty prevailed, and the careful use of language became crucial. In Stiglitz’s view, the IMF’s wording amounted to shouting “fire” in a crowded theatre: the rhetorical equivalent of triggering panic from within. This spectacular dissolution also corresponds to what later came to be called the “Minsky moment,” marked by uncertainty and distress preceding the crisis. It is also in this precise moment that language carries maximum weight — when suggestion fully exerts its power. Minsky (1986) called on central banks to intervene rapidly as lenders of last resort, that is, to assume the role of *leader*. The objective was to put out the metaphorical fire, to calm the crowd. Although he articulated concrete policy recommendations, his argument implicitly recognised the symbolic function of the lender of last resort as a stabilising presence — much like the leader in crowd psychology.

When, after the outbreak of the *subprime* crisis, the Federal Reserve announced a massive injection of liquidity, the flames in financial markets began to recede. Although belatedly, when the European Central Bank followed the Fed’s example and announced a rescue plan of comparable magnitude — with an unlimited amount and an intentionally unspecified duration — both interbank and sovereign rates stabilised. Mario Draghi’s now-famous “whatever it takes” (2012) is a textbook illustration of the leader’s capacity to *suggestionner* a psychological crowd, and thereby to reduce noise and create news in order to re-anchor expectations. The link between central bank communication and the calming of panic — particularly when the communication is delivered by the president himself — has been empirically established more in terms of *volume* of communication than its substantive economic content (Istrefi, Odendahl and Sesteieri, 2021). In Europe, however, what remained most vividly in collective memory was Draghi’s “whatever it takes,” which also earned him the nickname “Super Mario” in the *Financial Times*, where the words were reinforced by a striking accompanying image (Steen, 2012, see figure to the side). This is precisely why the *discourse channel* of central banks is so important: it is through this channel that the institution assumes the psychological function of the *leader*, the only agent capable of suggestionning the crowd and exercising the kind of viral power described in epidemic models, thereby restoring direction. As Flynn and Sastry (2025, p. 55) remark, "the narratives introduced by policymakers have the potential for significant impact. We know relatively little about what makes a policy narrative into a great story: Why, for example, did Mario Draghi's unscripted remarks about doing 'whatever it takes' make a much more compelling story than similar statements by other central banks?" Mass psychology can answer this question: it provides the keys to communicating with a crowd — through images, powerful words, and striking formulas. "A happy expression, an apt image, has sometimes diverted a crowd from the most sanguinary acts" … In three words: "Strike the imagination" (Le Bon, op. cit., p. 19). An in-depth analysis of the role of the leader will, however, be the object of a subsequent article.

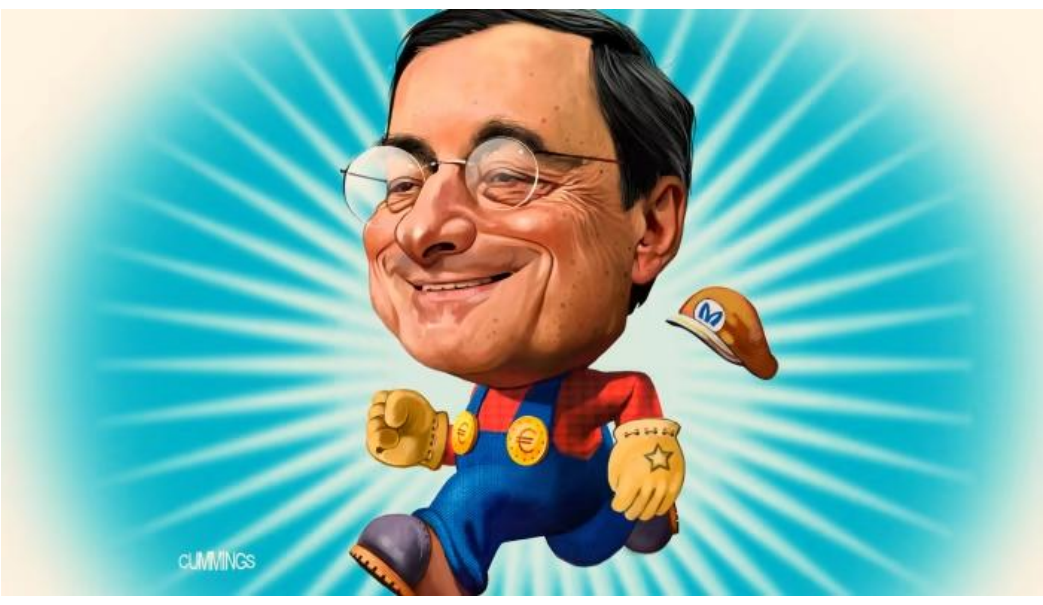

Finally, in his dissertation, Hansen (op. cit.) discusses the work of Jones (1900), "assistant professor of Economics and Commercial Geography of the University of Michigan Edward [who] drew on a range of ideas from crowd and imitation-suggestion theory in an attempt to carve out an outline of the 'chief psychological phenomena of crises'" (in Hansen, 2017, p. 84).

### *Contagion of sentiments and ideas in a single direction*

According to Le Bon (op. cit., p. 17), "thousands of otherwise separate individuals may, under the influence of intense collective emotions, suddenly acquire the characteristics of a psychological crowd. It is only at this advanced stage of organisation that new and special traits emerge, superimposed upon the underlying substrate, and these traits orient all sentiments and thoughts in a single direction". Le Bon refers to this as the "psychological law of the mental unity of crowds."

This is precisely the power of financial and economic narratives, and this is their consequence. As Shiller (2000, p. 497) concludes, "a great deal of evidence is presented here that suggests that social movements, fashions, or fads are likely to be important or even the dominant cause of speculative asset price movements." In finance, during conditions favorable to bubbles and crashes, it is through suggestion that the breach opens for informational cascades and emotional contagion. Shiller explicitly likens such narratives to viruses studied in epidemiology, insofar as they spread with exponential speed, like an epidemic — and often like a pandemic. A parallel can also be drawn with the fabulous, mythical, or mystical narratives described by Mackay (1852), whose "extraordinary delusions" generated genuine collective hysterias — precisely the kind of "mania" later analysed by Kindleberger. Freud (1921) likewise describes the crowd as resembling a personality marked by bipolar oscillation, shifting rapidly and dramatically from euphoric hyperactivity to systemic despair. This depiction mirrors the speed and amplitude with which a crowd can move, in the same way informational cascades unfold. In the words of Flynn and Sastry (2025), "sufficiently contagious narratives that cross a virality threshold can induce a phenomenon we call narrative hysteresis, in which one-time shocks can move the economy into stable self-fulfilling periods of optimism or pessimism" (p. 55). These forms of "narrative hysteresis" generate disproportionate enthusiasm — far beyond what any notion of rational assessment would allow.

Bikhchandani, Hirshleifer, and Welch (1992, p. 25) confirm that these cascades — which give rise to herding — can be triggered by "a small amount of information." This is entirely consistent with Le Bon's observation that, within a crowd, every sentiment and every act becomes contagious, spreading rapidly through mere psychological contact.
Studies of long-distance emotional contagion — notably on Facebook (Kramer et al., op. cit.), on Twitter (Ferrara and Yang, op. cit.), on financial discussion forums (Antweiler and Frank, op. cit.), and on crowdsourced forecasting platforms such as Estimize (Garcia, op. cit.) — show that contagion is often conveyed through nothing more than isolated words, chosen for their positive or negative emotional valence. These messages are characteristically short, non-technical, and often undocumented, yet exert a strong suggestive force precisely because their symbolic power generates vivid imagery in the reader's imagination. For the same reason, Stage (op. cit.) describes the audience of the blog written by Eva Dien Brine Markvoort as an "online crowd." She identifies Eva as a crowd leader in the Le Bonian sense, not because she argued or persuaded rationally, but because she suggestioned her audience through sheer affective resonance — in this case through a single video composed exclusively of facial expressions of suffering, without a single spoken word.

Emotional contagion, which is amplified within a crowd, has been extensively demonstrated. In their review, Herrando and Constantinides (2021) structure its effects into three domains: behavioural, physiological, and neurological. At the behavioural level, "reading reviews and observing other people's behavior on social networks have been proven to trigger emotional contagion. (…) These studies have confirmed that even in the complete absence of nonverbal cues, emotions can be contagious" (p. 3). At the physiological level, "over the years, social neuroscientists have provided evidence that, during social interactions, the observation of another person's emotional state automatically activates the same autonomic nervous system response and neural representation of the affective state as that of the observer" (p. 4). Finally, at the neurological level, "emotional processing can be monitored using neurophysiological tools and neuroimaging tools, since the level of arousal can be associated with specific brain activity in the prefrontal cortex (…). For example, functional magnetic resonance imaging (fMRI) has been used for the study of basic emotions such as happiness, sadness, fear, anger, disgust, and surprise (…) Functional near-infrared spectroscopy (fNIRS) has been used to monitor the role of the prefrontal cortex in emotion processing. (…) Electroencephalography (EEG) has been used to study affective states in the decision-making process (…). Due to the high temporal resolution of EEG, this neuroimaging tool has been widely employed to identify emotions, as an alternative to fNIRS or fMRI" (p. 4).
Hatfield, Cacioppo and Rapson (1994) show that emotions can be transmitted automatically within a group, even in the absence of verbal communication. Hatfield (1995) further notes that "child psychologists have collected some evidence that, from the start, both parents and children are powerfully 'enmeshed'; both parents and children show evidence of emotional contagion" (Thompson, 1987). Hatfield and her co-authors refer in particular to the experiment by Signer (1971), in which young children, aged between two and four, began to cry when they heard other children crying, yet did not react when the crying was "synthetic," that is, fabricated. Emotional contagion, inherent to human beings, is therefore magnified within a crowd.
Le Bon (op. cit., p. 29) makes this mechanism explicit: "the emotions expressed by a crowd, whether positive or negative, possess two fundamental qualities: they are extremely simple and they are extremely exaggerated. (…) Within the crowd, the exaggeration of a feeling is intensified because, spreading rapidly through suggestion and contagion, the approval it attracts greatly amplifies its force." This corresponds directly to the underreaction and overreaction phenomena in financial markets documented by behavioral finance.

De Bondt and Thaler (1990) also demonstrate that in finance no one escapes this contagious collective « effluve » — not even experts. Lo and Repin (op. cit.) observed the same pattern physiologically among ten traders during what they refer to as "transient market events," precisely the kind of circumstances that generate Le Bon's psychological crowd.

Thus, herding — as a crowd phenomenon resulting from suggestion, which itself gives rise to informational cascades and emotional contagion — is not merely a cognitive bias (Almansour, op. cit.) but an intrinsic property of mass behaviour. It is a characteristic inherent to living beings and one that becomes particularly amplified within the crowd.

***The tendency to transform suggested ideas immediately into action***

Le Bon notes that "for those who are under the influence of suggestion, the fixed idea tends to convert itself into action" (1895, p. 24). In this sense, suggestion and emotional contagion do not merely guide cognition; they guide behaviour. The issue here is not to restate the existence of herd-like reactions — these are empirically well-established — but to reinterpret them through the lens of *animal spirits*. We remind that Keynes (op. cit., p. 161) defined animal spirits as "a spontaneous urge to action rather than inaction," an impulse that drives not only investment but also speculation, thereby fuelling booms and, ultimately, busts. The inherently *gregarious* character of animal spirits has been widely documented (Akerlof and Shiller, 2009; Dow and Dow, 2011; Flynn and Sastry, 2025, among others). At the macroeconomic level, their aggregate impact has been quantified: Flynn and Sastry (2025) estimate "that narratives explain about 32 percent of the early 2000s recession and 18 percent of the Great Recession of 2008–09" (p. 55).

Nevertheless, following Dow and Dow (2011), our objective here is to focus on the strict definition of animal spirits — not as herding per se, but as a spontaneous and irrepressible *urge to act* — exactly as Le Bon conceptualises the instinctive nature of the crowd. As he writes, "by the mere fact that he forms part of a crowd, man descends several rungs on the ladder of civilisation. In isolation he may have been a cultivated individual; in the crowd he becomes an instinct-driven being, and therefore a primitive one. He possesses the spontaneity, the violence, the ferocity — and also the enthusiasms and heroism — of primitive beings" (Le Bon, 1895, p. 20). The crowd does not deliberate; it acts. Le Bon makes this even more explicit: "it need hardly be added that the inability of crowds to reason correctly deprives them of all critical spirit — that is, of the capacity to distinguish truth from error and to formulate a precise judgment" (op. cit., p. 36).

This aligns closely with the psychological notion of "Thought-Action Fusion" (TAF), which attributes excessive significance to thought and thereby precipitates impulsive action; this concept has been instrumental in advancing the understanding of obsessive compulsive disorder (Rachman, 1993). From a psychoanalytic perspective (Freud, 1921), the crowd may once again be likened to a bipolar individual, oscillating between manic phases — what Kindleberger (1978) describes as speculative "manias," marked by overconfidence and hyperactivity — and depressive phases characterised by defeatism and withdrawal. In Keynes's (1936) sense, animal spirits correspond to this manic state: a spontaneous urge to act that becomes almost uncontrollable. This "thought-into-action" mechanism, which follows the "thought-as-image" dynamic identified by Moscovici (1985), can also be linked to overconfidence bias, which fuels excessive risk-taking. In Keynes's framework, animal spirits are the psychological engine of optimism. In crash conditions, however, it is a more primitive survival instinct that drives the decision to sell. The instinct is undoubtedly "animalistic," yet the animal spirits in Keynes's original formulation refer above all to a willingness to act despite uncertainty — more akin to audacity, which easily slips into impulsivity.
The impulsive nature of financial markets has become increasingly well-documented since overconfidence bias entered the field of behavioural finance. Chhabra (2021) recently found "that impulsivity and risk level are positively correlated," even though non-impulsive participants do exist. The volume of suggestive messages on forums (Antweiler and Frank, op. cit.) predictably increases trading volume; but during a bubble — given the massive scale of aggregate buying — impulsivity appears not as an anomaly but as the norm, driven by animal spirits and exacerbated by regret aversion (Shefrin and Statman, op. cit.). This mechanism is also consistent with the findings of Au et al. (op. cit.), who show that a "good mood" leads to less precise decision making.

**Table 3: Behavioral Finance and Crowd psychology: A complementarity**

| | |
|---|---|
| Predominance of the unconscious personality | System 1 (Stanovich and Weist, 2000; Kahneman, 2002) |
| Orientation through suggestion | Stories telling (Shiller, 2000) and Attentional cascade (Shiller, 2000) |
| Affective contagion | Informational cascades (Scharfstein and Stein, 1990) and Emotional contagion (DeLong et al., 1990) |
| Immediate conversion of ideas into action | Animal spirits (Keynes, 1936) and Herding (Banerjee, 1992) |

Source: Author

Finally, a further explanatory element behind this defiant boldness in euphoric phases lies in the fact that the crowd is animated by extreme — and often contradictory — emotions. In this sense, the crowd does not merely *ignore* cognitive dissonance (Akerlof and Dickens, 1982); it *transcends* it. As Le Bon explains, "the elementary reasoning of crowds is, like higher reasoning, based on associations; but the associations formed by crowds are connected only by superficial resemblance or by simple succession. They link ideas in the manner of an Eskimo who, observing that ice — a transparent substance — melts in the mouth, concludes that glass, also a transparent substance, must likewise melt in the mouth. (…) The association of dissimilar things on the basis of merely apparent resemblance, and the immediate generalisation of particular cases — these are the defining characteristics of collective logic" (Le Bon, 1895, p. 37). Thus, even in the presence of alarming or contradictory information — information that, at an individual level, should induce cognitive dissonance and require a period of reflection to be integrated — the crowd appears simply to sweep it aside. In the context of a speculative bubble, despite signals suggesting that a reversal may already be underway, the market remains driven by the impulse to continue buying.

## Conclusion and implications for future research

Ultimately, Flynn and Sastry (op. cit., p. 52) also conclude that “viral narratives could be the missing link between emotions and economic fluctuations (…) Storytelling is central to how we interpret economic events. We recall economic history through haunting images of anxious crowds waiting to take money out of banks during the Great Depression or dejected office workers carrying cardboard boxes out of Lehman Brothers in 2008.” In the field of monetary policy, Chakraborty (2024, p. 1) likewise emphasises “the importance of incorporating behavioral insights to achieve a more nuanced understanding of economic dynamics and central bank communication.”
Le Bon (1895), Tarde (1901), and Freud (1921) indeed developed a predominantly pessimistic view of crowds — but only of crowds *abandoned to themselves*, that is, without a leader. All three liken the crowd to a primitive being, or even to a child. Freud compares it to an undisciplined child. These early theorists of crowd psychology were writing in a historical context in which crowd phenomena were frequently violent, sometimes even bloodthirsty. Fortunately, financial markets are not ordinarily destructive in such a literal sense. This article concerns the specific circumstances in which they *can* become so: when they form a heterogeneous and unregulated psychological crowd — that is, during periods of boom and bust.

Under such circumstances, markets have repeatedly demonstrated their capacity for impulsive and violent movements, producing severe macroeconomic and systemic consequences. In other words, left to themselves in moments of collective agitation, financial markets possess the capacity to generate crises with long-lasting repercussions, as the 2008 crisis so clearly illustrated. Yet as Ülgen (2021) reminds us, financial stability — a core component of macroeconomic stability — is a *public good*. Such a good cannot responsibly be left to the forces of crowd psychology, particularly since boom-and-bust phases are not rare exceptions, but recurrent features of financial capitalism.

Another research avenue would be to reinterpret the cognitive biases documented in behavioural finance — at the microeconomic and micropsychological level — through the analytical lenses of crowd psychology. This would offer a further perspective on the psychological mechanisms at work in financial markets and contribute to an even more comprehensive understanding of their dynamics.
A second avenue, directly inspired by the framework developed here, would be to address the question raised by Flynn and Sastry (2025) concerning the communication tools available to the regulator. In line with the analysis presented above, the regulator must assume the role of the *leader* — the only agent capable of orienting the crowd at the critical moment when it forms. This will be the subject of a subsequent article.

Because, in the end, as Le Bon warned, “crowds are like the Sphinx of ancient fable: either one learns to decipher the problems posed by their psychology, or one must resign oneself to being devoured by them” (1895, p. 55).